\documentclass[aps,pra,groupedaddress,twocolumn,longbibliography]{revtex4-1}
\usepackage[utf8]{inputenc}
\usepackage{amsmath}
\usepackage{amsfonts}
\usepackage{amssymb}
\usepackage{graphicx}
\usepackage{hyperref}
\usepackage{color}
\usepackage[T1]{fontenc}

\newcommand{\cre}[1]{\hat{#1}^{\dagger}}

\begin{document}


\title{Phonon signatures in spectra of exciton polaritons in transition metal dichalcogenides }

\author{F.~Lengers}
\affiliation{Institut f\"ur Festk\"orpertheorie, Universit\"at M\"unster,
Wilhelm-Klemm-Str.~10, 48149 M\"unster, Germany}

\author{T.~Kuhn}
\affiliation{Institut f\"ur Festk\"orpertheorie, Universit\"at M\"unster,
Wilhelm-Klemm-Str.~10, 48149 M\"unster, Germany}

\author{D.~E.~Reiter}
\affiliation{Institut f\"ur Festk\"orpertheorie, Universit\"at M\"unster,
Wilhelm-Klemm-Str.~10, 48149 M\"unster, Germany}

\begin{abstract}
Embedding a monolayer of a transition metal dichalcogenide in a high-Q optical cavity results in the formation of distinct exciton polariton modes. The polaritons are affected by the strong exciton-phonon interaction in the monolayer. We use a time convolutionless master equation to calculate the phonon influence on the spectra of the polaritons. We discuss the non-trivial dependence of the line shapes of both branches on temperature and detuning. The peculiar polariton dispersion relation results in a linewidth of the lower polariton being largely independent of the coupling to acoustic phonons. For the upper polariton, acoustic phonons lead to a low-energy shoulder of the resonance in the linear response. Furthermore, we analyze the influence of inhomogeneous broadening being the dominant contribution to the lower polariton linewidth at low temperatures. Our results point towards interesting phonon features in polariton spectra in transition metal dichalcogenides.
\end{abstract}

\maketitle
\section{Introduction}
The discovery of monolayer transition metal dichalcogenides (TMDCs) as atomically thin direct semiconductors \cite{Mak10,Splendiani10} has led to an extensive investigation of their physical properties in order to construct next-generation optoelectronic devices \cite{Mueller18,Autere18}. The most striking feature of TMDC monolayers is the exceptionally strong exciton binding energy of few hundreds of meV \cite{Chernikov14,Drueppel18}. If a TMDC monolayer is integrated into a photonic cavity such that a confined light mode interacts with the excitonic system new quasiparticles, the polaritons, emerge \cite{Dufferwiel15,Lundt16,Flatten16,Juergens20}.
Being mixtures of photons and excitons, polaritons gain the low effective mass of the photonic field and the non-linearities arising from the excitonic system. Those two features makes them promising for lasing applications and the investigation of condensation phenomena \cite{Schneider18,Kasprzak06}. Reaching the strong-coupling regime of TMDC monolayers within photonic cavities triggered research about polariton transport and condensation \cite{Schneider18,Waldherr18,Lundt19}.

The TMDC monolayers also feature an exceptionally strong exciton-phonon interaction, visible for example in phonon sidebands \cite{rosati2020temporal} or in pronounced asymmetric line shapes due to phonon-assisted processes \cite{Christiansen17,Shree18,Lengers20}. As the polaritons are composed of TMDC excitons, it is interesting to ask how phonons influence the polariton modes. Experimental results show that the linewidths of upper and lower polariton mode strongly depend on the detuning between exciton and photon \cite{Dufferwiel15}, while the specifics of linewidth and line shape and especially their origin still have to be clarified.

\begin{figure}[t!]
\includegraphics[width=\columnwidth]{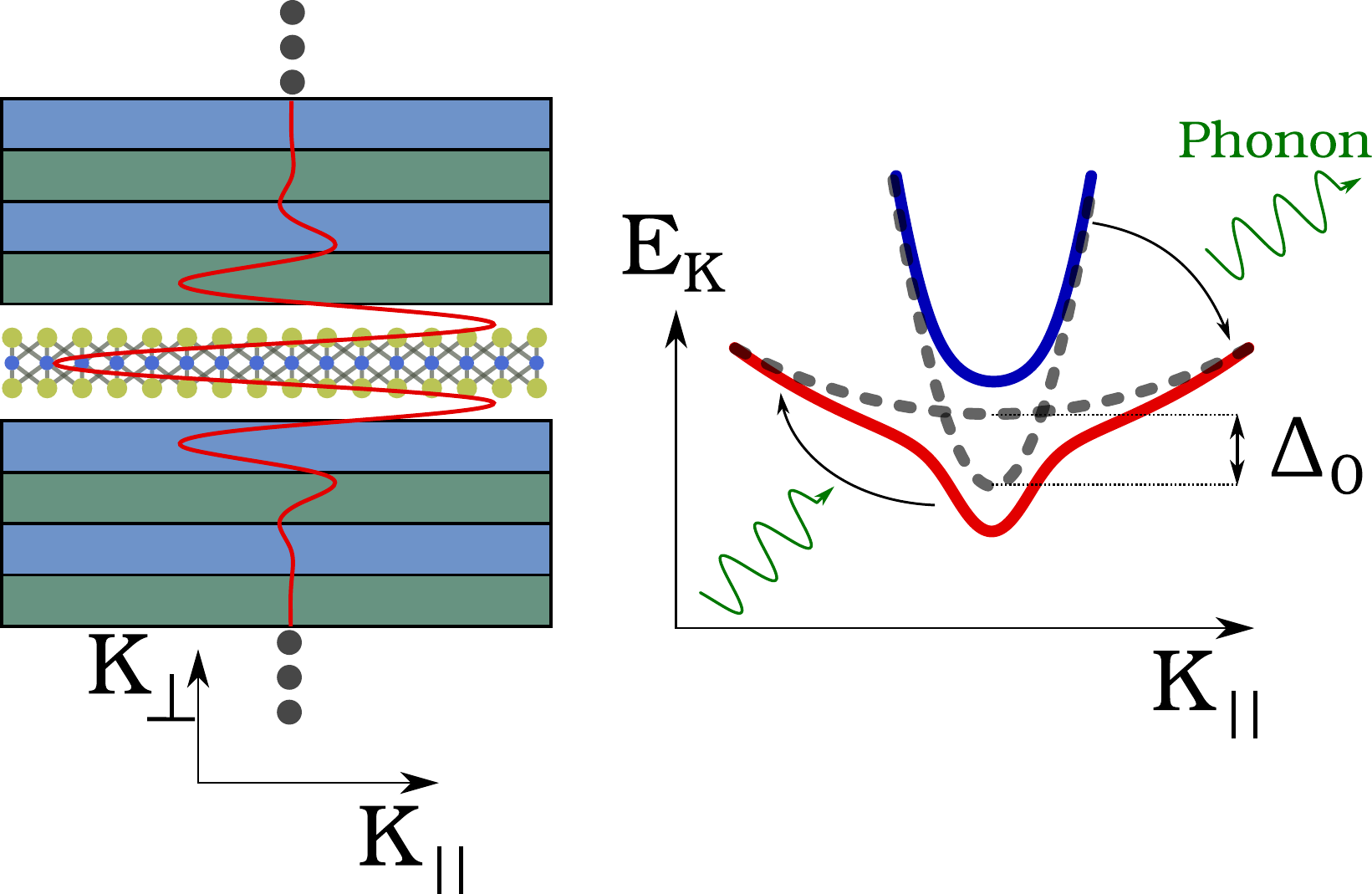}
\caption{TMDC monolayer within a photonic microcavity. The light-field is quantized in $\mathbf{K}_{\perp}$ direction such that only $\mathbf{K}_{\parallel}$ is a good quantum number and only one photonic mode with detuning $\Delta_0$ interacts with the excitonic dispersion forming a lower (red) and upper (blue) polariton mode. The exciton-phonon interaction translates to polariton-phonon interaction (green).}
\label{Fig_sketch}
\end{figure}

This paper is therefore devoted to discussing the interplay between polaritons and phonons. To this end, we calculate the linear optical response of a TMDC monolayer coupled to a photonic cavity under the influence of phonons and analyze the resulting line shapes with and without an additional inhomogeneous broadening. A sketch of the system is shown in Fig.~\ref{Fig_sketch} displaying the TMDC monolayer in a photonic cavity formed by two Bragg mirrors as well as the polariton dispersion relation including possible phonon scattering processes.

For our calculations, we apply a time convolutionless master equation for the polariton amplitudes to obtain the linear response in the low-density regime. This method has been shown to describe the linear absorption spectra including the phonon sidebands of a bare TMDC monolayer in a compact way \cite{Lengers20}. By analyzing a wide temperature range, discussing the influence of phonons, detuning and inhomogeneous broadening, we provide a thorough understanding of the polariton-phonon coupling. 

\section{Theory}
\subsection{Hamiltonian in the exciton basis}
The Hamiltonian describing the system of a TMDC monolayer in a photonic microcavity as sketched in Fig.~\ref{Fig_sketch} is given by

\begin{align}
\hat{H}
=&
\sum\limits_{\mathbf{K}}E_{\mathbf{K}}\cre{X}_{\mathbf{K}}\hat{X}_{\mathbf{K}}+
\sum\limits_{\mathbf{K}}\hbar\Omega_{\mathbf{K}}\cre{a}_{\mathbf{K}}\hat{a}_{\mathbf{K}}
\label{Eq_H} \\
+&g\sum\limits_{\mathbf{K}}\left(\cre{X}_{\mathbf{K}}\hat{a}_{\mathbf{K}}+\cre{a}_{\mathbf{K}}\hat{X}_{\mathbf{K}}\right)
\nonumber\\
+&\sum\limits_{\mathbf{K},i}\hbar\omega_{\mathbf{K},i}\cre{b}_{\mathbf{K},i}\hat{b}_{\mathbf{K},i} \nonumber \\
 +&\sum\limits_{\mathbf{K},\mathbf{Q},i}g_{\mathbf{Q},i}\cre{X}_{\mathbf{K+Q}}\hat{X}_{\mathbf{K}}
 \left(\hat{b}_{\mathbf{Q},i}+\cre{b}_{-\mathbf{Q},i}\right) .
\nonumber
\end{align}

The first two terms of Eq.~\eqref{Eq_H} describe the free exciton and photon system with the creation (annihilation) operators for excitons $\cre{X}_{\mathbf{K}}$ ($\hat{X}_{\mathbf{K}}$) and photons $\cre{a}_{\mathbf{K}}$ ($\hat{a}_{\mathbf{K}}$) with in-plane momentum $\mathbf{K}$.

The exciton energy is given in effective mass approximation by 
	$$E_{\mathbf{K}}= E_{1s}+\frac{\hbar^2K^2}{2M}$$
with the total exciton mass $M$ and the bound-exciton energy $E_{1s}$. We focus on $\mathrm{MoSe_2}$ as material, such that the bright A excitons within the K and K' valleys are the energetically lowest (1s) excitonic states \cite{Malic18}, which can be selectively addressed using circularly polarized light \cite{Xiao12,Wang18}. Considering quasi-resonant coupling of the $1s$ exciton to the cavity mode, we also neglect intervalley coupling by a combination of exciton-phonon and exciton-exchange interaction \cite{Moody15,Kioseoglou16,Schmidt16,Selig19} as an approximation, because we do not focus on valley polarization. We further use $M=1.1m_0$ with $m_0$ being the free electron mass.

The photon dispersion is given by $\hbar\Omega_{\mathbf{K}}$. We take into account one subband of light confined by perfectly reflecting mirrors leading to a quantization of the out-of-plane momentum $K_z$ orthogonal to the TMDC plane, yielding
\begin{align*}
	\hbar\Omega_{\mathbf{K}}&=\hbar c\sqrt{K_z^2+\mathbf{K}^2} \, \nonumber \\
		& = E_{\text{cav}} \sqrt{1+\mathbf{K}^2/K_z^2} \nonumber
\end{align*}
with $c$ being the speed of light in vacuum and $E_{\text{cav}}= \hbar c K_z$. We assume all other subbands to be strongly off-resonant to the excitonic system such that the usual requirements for exciton-polaritons are fulfilled. 

The third term in Eq.~\eqref{Eq_H} describes the exciton-photon interaction in rotating-wave approximation with a constant interaction strength $g$ between the in-plane exciton and one in-plane photon branch confined in the out-of-plane direction by the cavity. Throughout this paper we fix the coupling strength to $g=15~$meV, which is a typical value in such systems \cite{Dufferwiel15,Lundt16}. Due to the rotating-wave approximation the detuning $\Delta_0$ between the exciton and photon mode at $\mathbf{K}=0$ 
 $$\Delta_0=   E_{1s} - E_{\text{cav}}$$
  is the relevant quantity for the dynamics instead of $E_{\mathrm{1s}}$ and $E_{\mathrm{cav}}$. Note that with our definition the lowest exciton state is below the photon dispersion for negative $\Delta_0$ and above for positive $\Delta_0$.

The last two terms in Eq.~\eqref{Eq_H} describe the free phonon system and the exciton-phonon interaction with the phonon operators $\cre{b}_{\mathbf{K},i}$ ($\hat{b}_{\mathbf{K},i}$).  The phonons of branch $i$  have the energy $\hbar\omega_{\mathbf{K},i}$. Here, we consider one effective optical phonon branch with energy $\hbar\omega_{\mathbf{K},\mathrm{opt}}=34.4~$meV and one effective acoustic phonon branch with $\hbar\omega_{\mathbf{K},\mathrm{ac}}=\hbar c_{\text{s}}K$ with the speed of sound $c_{\text{s}}=4.1~\mathrm{nm\cdot ps^{-1}}$ . The exciton-phonon matrix elements $g_{\mathbf{Q},i}$ are determined in deformation potential approximation using parameters taken from ab-initio calculations \cite{Jin14} and the specific values for $\mathrm{MoSe_2}$ can be found in Ref.~\cite{Lengers20}. To determine the $1s$ exciton wave function entering the exciton-phonon interaction, we consider a monolayer of $\mathrm{MoSe_2}$ on a $\mathrm{SiO_2}$ substrate formed by the Coulomb interaction modeled by the Rytova-Keldysh potential \cite{Berghaeuser14}. While different dielectric surroundings have a profound impact on the excitonic properties, especially the exciton binding energy \cite{Drueppel17,Waldecker19}, we take the absolute energetic position of the $1s$ exciton as a parameter and focus on the qualitative features of polariton-phonon interaction which should be captured by the applied model as long as the $1s$ exciton is sufficiently separated from other excitonic resonances.

\subsection{Hamiltonian in the polariton basis}
We next transform the given Hamiltonian into the polariton basis by diagonalizing the coupled exciton-photon system. This is readily done, because the in-plane momentum $\mathbf{K}$ is a good quantum number. This results in the formation of the polariton creation and annihilation operators  $\cre{P}_{\mathbf{K},\Lambda}$ and $\hat{P}_{\mathbf{K},\Lambda}$ given by 
	\begin{align}
	\cre{P}_{\mathbf{K},\Lambda}=
	&\alpha_{\mathbf{K},\Lambda}\cre{X}_{\mathbf{K}}+ \beta_{\mathbf{K},\Lambda}\cre{a}_{\mathbf{K}},
	\label{Eq_Polariton_Basis}
	\end{align}
	where $\alpha_{\mathbf{K},\Lambda}$, $\beta_{\mathbf{K},\Lambda}$ are the exciton and photon contribution to the polariton mode $\Lambda$, respectively. The energies of the polariton modes are
	$$\mathcal{E}_{\mathbf{K},\Lambda}=\frac{E_{\mathbf{K}}+\hbar\Omega_{\mathbf{K}}}{2}\pm\sqrt{\left(\frac{E_{\mathbf{K}}-\hbar\Omega_{\mathbf{K}}}{2}\right)^2+|g|^2},$$
where we see that an upper polariton ($\Lambda=$UP,"$+$") and lower polariton ($\Lambda=$LP,"$-$") mode emerges. The corresponding energy dispersions are sketched in Fig.~\ref{Fig_sketch}. 
For the uncoupled case we have two parabolic dispersions for the excitons and photons (grey dashed lines). Due to the interaction an anti-crossing occurs and the LP (red) and UP (blue) branches form. The composition of the polaritons given by the coefficients $\alpha_{\mathbf{K},\Lambda} $ and $\beta_{\mathbf{K},\Lambda}$ depends sensitively on the detuning $\Delta_0$, which is sketched for the photon contribution $|\beta|^2$ for the case $\mathbf{K}=0$ in Fig.~\ref{Fig_sketch_spectrum}(left).

For a detuning $\Delta_0=0$, both polaritons at $\mathbf{K}=0$ consist of $50\%$ exciton and photon. For positive detunings $\Delta_0>0$ the LP is dominantly photon-like and consequently the exciton contribution decreases. The opposite is true for $\Delta_0<0$, where the UP is photon-like.

In the polariton basis the Hamiltonian reads 
	\begin{align}
	\hat{H}=&\sum\limits_{\mathbf{K},\Lambda}\mathcal{E}_{\mathbf{K},\Lambda}\cre{P}_{\mathbf{K},\Lambda}\hat{P}_{\mathbf{K},\Lambda}
	\label{Eq_H_PolaritonBasis}\\
	+&\sum\limits_{\mathbf{K}}\hbar\omega_{\mathbf{K}}\cre{b}_{\mathbf{K}}\hat{b}_{\mathbf{K}} \nonumber \\
	+&  \sum\limits_{\mathbf{K,Q},\Lambda,\Lambda',i} g_{\mathbf{Q},\mathbf{K},i}^{\Lambda\Lambda'}\cre{P}_{\mathbf{K+Q},\Lambda}\hat{P}_{\mathbf{K},\Lambda'}\left(\hat{b}_{\mathbf{Q},i}+\cre{b}_{-\mathbf{Q},i}\right) \,.
	\nonumber
	\end{align}
The last line  in Eq.~\eqref{Eq_H_PolaritonBasis} describes the effective polariton-phonon coupling with the transformed coupling matrix element
	$$g_{\mathbf{Q},\mathbf{K},i}^{\Lambda\Lambda'}=g_{\mathbf{Q},i}\left(\alpha_{\mathbf{K+Q},\Lambda}^*\alpha_{\mathbf{K},\Lambda'}\right).$$
Phonons only couple to the excitonic contribution of the polariton branches and therefore the polariton-phonon coupling $g^{\Lambda\Lambda'}$ is proportional to the coefficient $\alpha$ of the respective polariton branch. The range of momenta in which photon and exciton are strongly coupled can be approximated by the width $K_0$ of the unconfined light cone affecting the exciton with $K_0 \lesssim \frac{E_{1s}}{\hbar c}\approx  0.01~\mathrm{nm^{-1}}$. Since typical polariton-splittings are in the order of $10-30~$meV \cite{Dufferwiel15,Lundt16,Flatten16}, phonon transitions between the two polariton branches can play a major role in their relaxation kinetics and optical spectra. We here focus on the case of low polariton density and linear spectra, where polariton-polariton interaction can be neglected. The latter becomes important in particular in the regime of polariton condensation \cite{Kasprzak06,Schneider18,Stepanov21}. 

\subsection{Calculation of the optical spectrum}

\begin{figure}[b]
\includegraphics[width=\columnwidth]{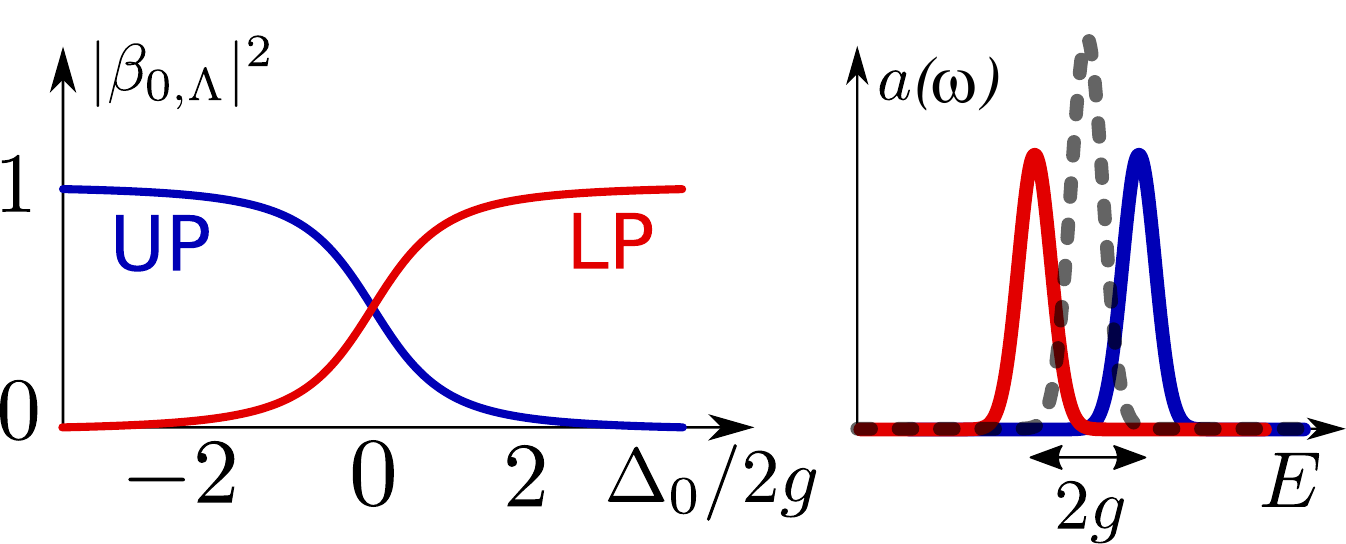}
\caption{Left: Photon amplitude $|\beta_{0,\Lambda}|^2$ (left) for $\mathbf{K}=0$ of the UP (blue) and LP (red) as a function of the detuning $\Delta_0$. Right: Sketch of a spectrum without phonons for $\Delta_0=0$ together with the spectrum of the bare cavity (grey dashed line).}
\label{Fig_sketch_spectrum}
\end{figure}

The quantity of interest of the polariton system is its optical spectrum. We assume that initially the cavity is occupied with a coherent light field with $\mathbf{K}=0$, which subsequently couples to the excitonic system, i.e., we take $a_0=\langle{\hat{a}_{\mathbf{0}}}\rangle(t=0)$ and $a_0\in \mathbb{R}$. This  leads to the initial condition for the polariton amplitudes 
	$$\langle \hat{P}_{\mathbf{0},\Lambda}\rangle(t=0)=\beta_{\mathbf{0},\Lambda}^* a_0.$$
Considering the linear response of the system, the spectrum is calculated as real part of the Fourier transform of the expectation value of the photon operator
\begin{eqnarray}
a(\omega) &=& \text{Re}\left[{\cal FT}\left(\langle{\hat{a}_{\mathbf{0}}}\rangle(t)\right)\right] \nonumber\\
	&=&\text{Re}\left[ {\cal FT}\left(\sum\limits_{\Lambda} \beta_{\mathbf{0},\Lambda}\langle \hat{P}_{\mathbf{0},\Lambda}\rangle(t)\right)\right]\,.
\end{eqnarray}
A spectrum without phonons in the case of vanishing detuning $\Delta_0=0$ is sketched in Fig.~\ref{Fig_sketch_spectrum}(right) for the uncoupled (grey dashed lines) and coupled (blue and red lines) system. The coupling of exciton and photon leads to a splitting into UP and LP, which are well visible in the spectrum if the splitting is larger than the linewidth. For a vanishing detuning $\Delta_0=0$ the two polariton peaks are at the energies $E_{\text{UP/LP}}=E_{\mathrm{cav}}\pm g$, i.e., their splitting is $2g$.

To calculate the phonon influence on the spectrum, we solve the equations of motion within a time convolutionless approximation (TCL) for the coherent polariton amplitude $P_{\mathbf{K},\Lambda}=\langle \hat{P}_{\mathbf{K},\Lambda}\rangle$ reading
	\begin{align}
	\frac{d}{dt}P_{\mathbf{K},\Lambda}=-\frac{i}{\hbar}\mathcal{E}_{\mathbf{K},\Lambda}P_{\mathbf{K},\Lambda}-\sum\limits_{\Lambda'}\Gamma_{\Lambda,\Lambda'}(\mathbf{K},t)P_{\mathbf{K},\Lambda'} \,.
	\label{Eq_EOM}
	\end{align}
The momentum- and time-dependent coupling $\Gamma_{\Lambda,\Lambda'}(\mathbf{K},t)$ is given by
	\begin{align}	\label{Eq_Gamma}
	&\Gamma_{\Lambda,\Lambda'}(\mathbf{K},t)
	=
	\frac{1}{\hbar^2}\sum\limits_{\mathbf{Q},\Lambda'',i,\pm} g_{\mathbf{Q},\mathbf{K-Q},i}^{\Lambda,\Lambda''}\left(g_{\mathbf{Q},\mathbf{K-Q},i}^{\Lambda'\Lambda''}\right)^*\\
	&\qquad \cdot \left(\frac{1}{2}\pm\frac{1}{2}+n_{\mathbf{Q},i}\right)\frac{e^{-\frac{i}{\hbar}\left(\mathcal{E}_{\mathbf{K-Q},\Lambda''}-\mathcal{E}_{\mathbf{K},\Lambda'}\pm\hbar\omega_{\mathbf{Q},i}\right)t}-1}{-\frac{i}{\hbar}\left(\mathcal{E}_{\mathbf{K-Q},\Lambda''}-\mathcal{E}_{\mathbf{K},\Lambda'}\pm\hbar\omega_{\mathbf{Q},i}\right)}
	\nonumber
	\end{align}
with the thermal incoherent phonon occupation $n_{\mathbf{Q},i}$ of mode $\mathbf{Q}$ and branch $i$.

The above equations of motion are derived by first setting up the equations of motion for the phonon-assisted polarizations $\langle\hat{P}_{\mathbf{K-Q},\Lambda}\hat{b}^{}_{\mathbf{Q},i}\rangle$ and $\langle\hat{P}_{\mathbf{K-Q},\Lambda}\hat{b}^{\dagger}_{\mathbf{-Q},i}\rangle$ in second Born approximation within a correlation expansion scheme \cite{Rossi02,Lengers20} and then adiabatically integrating the equations of motion in Born approximation using an initial time $t_0=0$ of optical excitation of the polariton modes at normal incidence by an ultrafast external source. Due to the time dependence of $\Gamma$ non-Markovian features like phonon sidebands are dynamically included in the description \cite{Lengers20}. The important steps of the derivation are given in App.~\ref{App_Derivation}. Such an approximate solution of the infinite hierarchy of equations of motion has been shown to yield accurate results when compared to high-order Born approximations in Ref.~\cite{Lengers20} in the case of a bare monolayer coupled to phonons. It should be noted that it is mandatory to formulate the TCL in a suitable basis, i.e., the polariton basis, to correctly account for the low-energy excitations of the system, while a general correlation expansion treatment without Markov approximation is independent of the chosen basis \cite{Siantidis01,Rossi02}.

In a real system, there are several dephasing channels apart from phonons. On the one hand, the confined cavity mode couples to unconfined continuum modes and thereby leaks out of the cavity, resulting in a the decay rate $\gamma_{\mathrm{cav}}$. Additionally, the exciton may decay radiatively into photonic continuum modes described by the a decay rate $\gamma_{\mathrm{ex}}$. These rates can be combined to calculate the decay rate of the polaritons  $\gamma_{\mathbf{K}}^{\Lambda\Lambda'}=\gamma_{\mathrm{cav}} \beta_{\mathbf{K},\Lambda}^*\beta_{\mathbf{K},\Lambda'}+\gamma_{\mathrm{ex}}\alpha_{\mathbf{K},\Lambda}^*\alpha_{\mathbf{K},\Lambda'}$. To simplify the discussion and to focus on the phonon-related effects we set $\gamma_{\mathrm{ex}}=\gamma_{\mathrm{cav}}=2\,\mathrm{ps^{-1}}$ resulting in the constant rate $\gamma_0$. Thereby the linewidth of the polariton resonances without phonons is independent of the respective exciton and photon contributions to the polariton mode. The equation of motion therefore is extended by
	\begin{align}
	\frac{d}{dt}P_{\mathbf{0},\Lambda}\Big|_{D}=-\frac{1}{2}\gamma_0 P_{\mathbf{0},\Lambda}.
	\label{Eq_Dissipator}
	\end{align}

After initializing the polariton polarizations, the resulting equations of motion of $P_{\mathbf{0},\Lambda}$ as a sum of Eq.~\eqref{Eq_EOM} and Eq.~\eqref{Eq_Dissipator} are propagated in time and $\Gamma_{\Lambda,\Lambda'}$ from Eq.~\eqref{Eq_Gamma} is calculated by separating the $K$-space in a polariton-region with $|\mathbf{K}|\le 2K_0$ and a region $|\mathbf{K}|\in\,]2K_0,5.5\,\mathrm{nm^{-1}}]$; in the latter case only the lower polariton is relevant for polariton-phonon interaction, because the UP has no exciton-contribution. Both of these regions are discretized on an equidistant grid of $400$ $K$-points.

We have given the equations of motion above for the general case of finite in-plane momentum $\mathbf{K}$ of the polariton modes, while due to excitation at normal incidence only the in-plane mode $\mathbf{K}=0$ is excited. We therefore drop the index $\mathbf{K}$ of all quantities in the following and $\mathbf{K}=0$ is always implied.

\section{Results}

\subsection{Temperature dependent polariton spectra}
\begin{figure}
\includegraphics[width=\columnwidth]{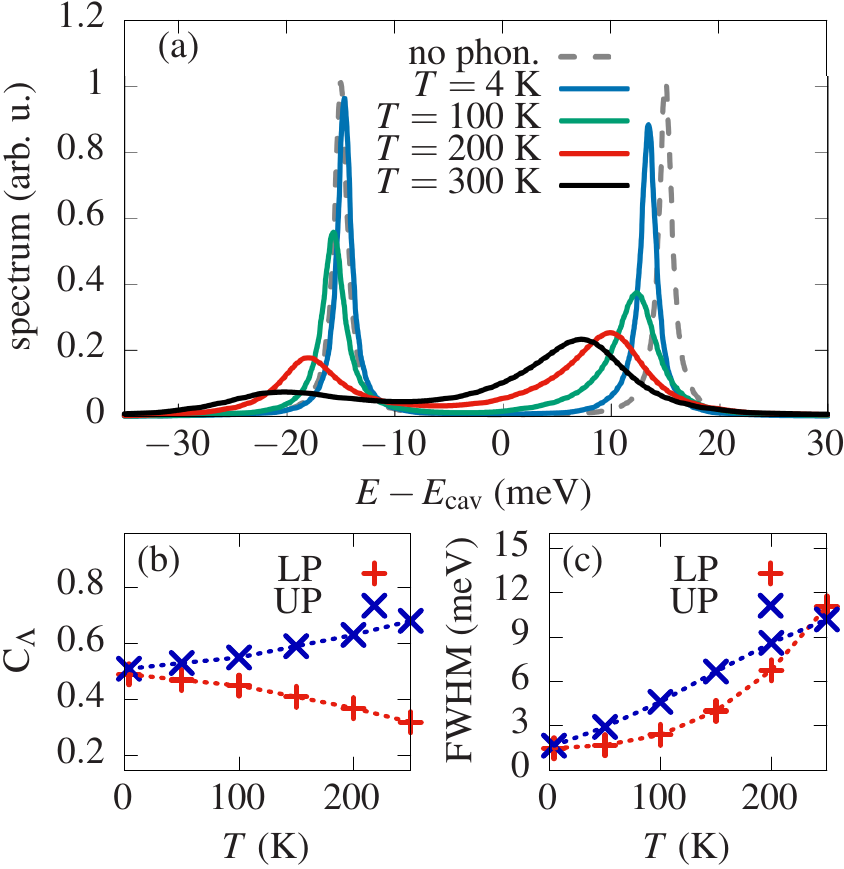}
\caption{(a) Spectrum as function of energy relative to the cavity energy $E_{\mathrm{cav}}$ for different temperatures, (b) fractional contributions of the LP (red) and UP (blue) to the spectrum and (c) FWHM of the respective resonances as function of temperature.
}
\label{Fig_Spectra_g15}
\end{figure}

Figure~\ref{Fig_Spectra_g15}(a) shows the spectra without phonons (grey dashed lined) and with phonons for four different temperatures at $T=4,100,200,300$~K. In the case without phonons we take $\Delta_0=0$, such that we obtain two peaks corresponding to the two polariton branches split by $2g=30$~meV and the linewidth is given by $\gamma_0$. 

When including phonons, there are several effects: On the one hand, the exciton-phonon interaction results in a shift of the exciton dispersion due to the formation of the polaron. To compensate for the low-temperature polaron shift of the exciton to lower energies of about $11~$meV \cite{Lengers20}, we set the single-particle detuning to $\Delta_0=11~$meV for better comparison. Therefore for $4$~K this corresponds to having no effective detuning. Nonetheless, a reduced total splitting compared to the phonon-free case can be observed due to a reduced effective exciton-photon coupling because the exciton is dressed by the phonon field which in turn does not couple directly to the light field \cite{Groll20}. 

Furthermore, the phonons affect the line shape and intensity. Already for $4$~K, the peak intensity of the UP is reduced compared to the case without phonons. This results from an increased broadening of the UP due to scattering by acoustic phonon emission into lower-energetic states. This process does not affect the LP because there are no lower-energetic states available which can be reached by phonon emission.

When increasing the temperature, we observe a gradual broadening and shift to lower energies of both polariton resonances.
For increasing temperature the broadening of the polariton lines increases due to increased phonon occupation. While for temperatures below $200~$K the LP exhibits the dominant peak intensity in Fig.~\ref{Fig_Spectra_g15}(a), for temperatures above $200~$K the UP dominates the spectra. This crossover can be explained by the temperature-dependent polaron shift of the system. As temperature increases, the polaron shifts the excitonic dispersion to lower energies \cite{Christiansen17,Lengers20}, therefore out of resonance with the photon field. While LP and UP have equal photon and exciton contributions at resonance, the LP is exciton-like for negative detunings. Thereby the LP couples more strongly to phonons for negative detunings (higher temperatures) and is also less visible in the cavity spectrum because it has a lower contribution of the photon field for negative detunings [cf. Fig.~\ref{Fig_sketch_spectrum}(left)]. 

These features are quantified in Fig.~\ref{Fig_Spectra_g15}(b) and (c) \footnote{Note that we do not include the results for $T=300~$K in Fig.~\ref{Fig_Spectra_g15}(b) and (c) because for this temperature LP and UP cannot be clearly distinguished [cf. Fig.~\ref{Fig_Spectra_g15}(a)].}. In Fig.~\ref{Fig_Spectra_g15}(b) we plot the fractions $C_{\Lambda}$ of the LP (red) and UP (blue) signals to the total signal
	$$C_{\Lambda}=\frac{\alpha_{\Lambda}}{\alpha_{\mathrm{UP}}+\alpha_{\mathrm{LP}}}$$
with $\alpha_{\Lambda}$ being the integrated spectrum around the $\Lambda$ resonance. We separate the two resonances by determining the energetic position of the minimum in the spectrum between the two peaks. The resulting contributions in Fig.~\ref{Fig_Spectra_g15}(b) clearly show that the contribution of the LP reduces with increasing temperature because of the increasing detuning. Only for $T=4~$K photon and exciton system are in resonance and it holds that  $C_{\Lambda}\approx 0.5$. 

In Fig.~\ref{Fig_Spectra_g15}(c) we plot the temperature-dependent full width at half maximum (FWHM) of the LP and UP. While at low temperatures the UP shows a larger FWHM due to the aforementioned possible relaxation channels to lower-energetic states, above $200~$K the FWHM of the LP dominates. This is attributed to the increase of exciton contribution of the LP as found in (a) and to an increased possibility of optical phonon absorption which may overcome the polariton binding energy. 

When comparing these results with experiments, it should be noted that the temperature-dependent detuning between photon and exciton is not fully described by the polaron shift discussed here. Instead the phenomenological Varshni shift in experiments is found to be larger than the polaron shift \cite{Arora15,Christiansen17}. Additionally the energy of the cavity mode can also slightly change as function of temperature \cite{reithmaier2004strong}. Such shifts may quantitatively modify the temperature dependence, but will result in a similar qualitative picture of the temperature dependence with the reduction of the LP signal as function of temperature.

\subsection{Dependence on the detuning}
\begin{figure}[b]
\includegraphics[width=\columnwidth]{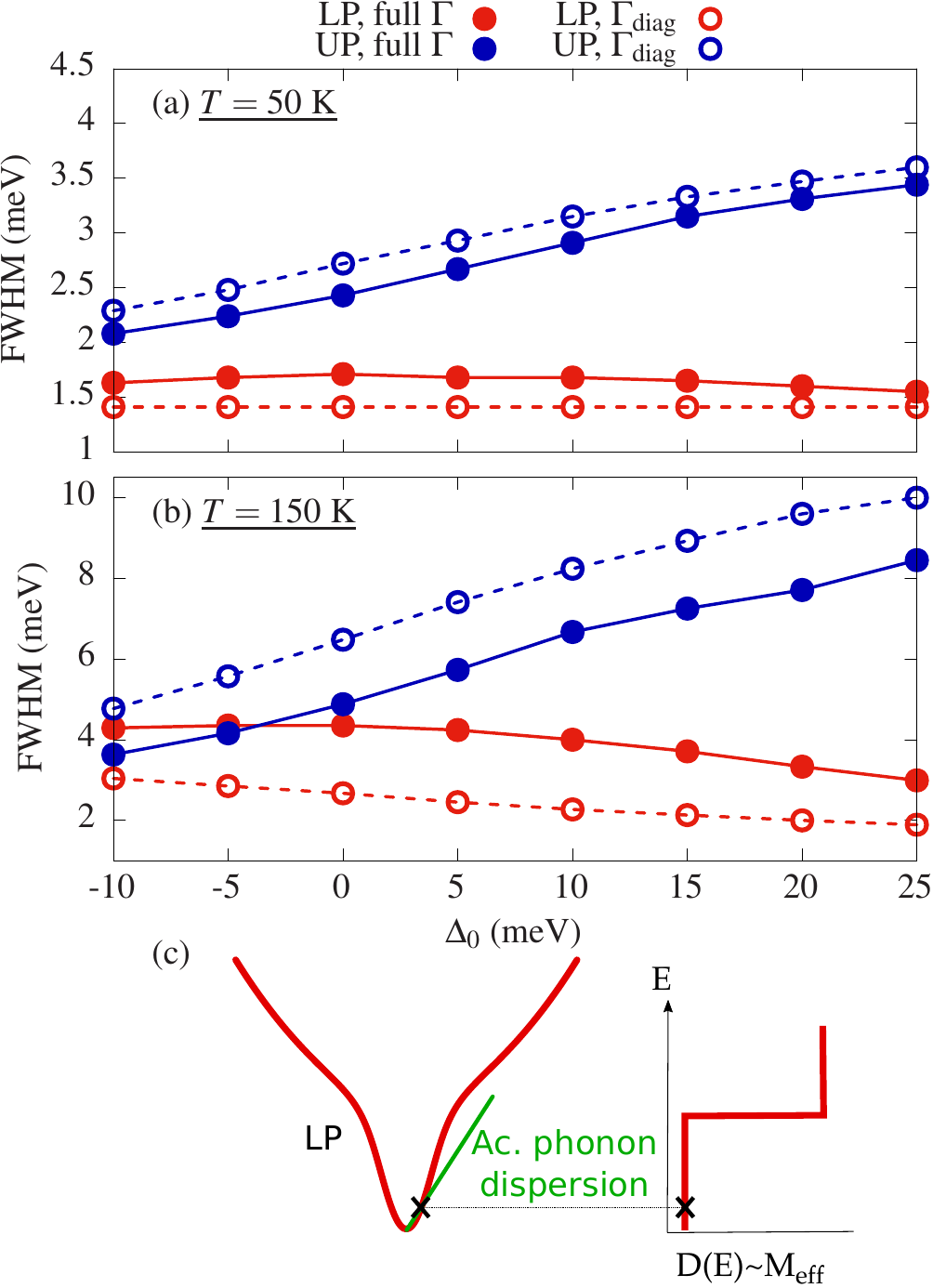}
\caption{FWHM of the UP (red) and LP (blue) peaks for (a) $T=50~$K and (b) $T=150~$K as function of the detuning $\Delta_0$. Filled circles (solid lines) correspond to the full calculation, empty circles (dashed lines) are without the off-diagonal elements of $\Gamma$. (c) Sketch of dephasing by acoustic phonon absorption in the case of the LP (left) together with a sketched density of states (right). The LP dispersion is sketched in red and acoustic phonon dispersion in green with the position of a resonant scattering by acoustic phonon absorption marked by a black cross.}
\label{Fig_det_specs_full}
\end{figure}

We have seen above that the temperature-dependent detuning of exciton and photon has a profound impact on the optical properties of the polariton spectra. To analyze the detuning dependence in detail, we fix the temperature to a certain value and vary the detuning $\Delta_0$. Experimentally, this can be realized by a tunable microcavity \cite{Dufferwiel15}. In Fig.~\ref{Fig_det_specs_full} we plot the FWHM of UP (blue) and LP (red) line as functions of detuning $\Delta_0$  for (a) $T=50~$K  and (b) $T=150~$K (b). The filled circles (solid lines) refer to the full calculation, while the empty circles (dashed lines) are calculated in an approximate solution as described below. 

For the UP (blue), we find that the FWHM increases monotonically as function of the detuning for both temperatures. This is reasonable because the excitonic component of the UP increases for increasing detuning and therefore couples more strongly to phonons. The opposite is true for the LP (red) such that for $T=150~$K (b) one observes a gradual decrease of linewidth for increasing detuning. However, the linewidth of the LP decreases not as strongly as the linewidth of the UP increases. For $T=50~$K in Fig.~\ref{Fig_det_specs_full}(a) the LP linewidth almost does not change at all. 

It is instructive to notice that the dephasing processes for UP and LP are not strictly independent of each other since the dephasing function $\Gamma_{\Lambda,\Lambda'}$ from Eq.~\eqref{Eq_Gamma} couples both polariton amplitudes for $\Lambda\neq\Lambda'$. Neglecting such non-diagonal contributions of $\Gamma$ results in independent dephasing processes for the two polaritons, which is often employed to quantify dephasing. In Fig.~\ref{Fig_det_specs_full} we plot the case when neglecting the non-diagonal contribution of $\Gamma$ (labeled as $\Gamma_{\mathrm{diag}}$) as empty circles (dashed lines). For both temperatures the non-diagonal contributions have a non-negligible impact on the linewidth and always lead to a decrease of the UP FWHM and an increase of the LP FWHM. Importantly, neglecting the non-diagonal contributions for $T=50~$K [cf. Fig.~\ref{Fig_det_specs_full}(a)] results in a constant FWHM of the LP as a function of detuning which is solely given by the decay rate $\gamma_0$. This illustrates that for $T=50~$K only the non-diagonal coupling of the two polariton branches results in a detuning-dependent linewidth of the LP. 

This behavior is linked to an efficient decoupling of the LP mode from acoustic phonons. Figure~\ref{Fig_det_specs_full}(c) shows a sketch of the LP dispersion together with the dispersion relation of acoustic phonons (left) as well as a sketch of the polaritonic density of states (right). A resonant dephasing process by acoustic phonon absorption is possible if the two lines cross (marked by a black cross) with an energy conserving scattering process determined by $E_{\mathrm{LP},\mathbf{K}} - E_{\mathrm{LP},0}-\hbar\omega_{\mathrm{ac},-\mathbf{K}}=0$. Due to the bending of the LP dispersion to lower energies, the relevant dispersion relation for this process acquires the effective mass of the photon state, which leads to a strongly decreased density of states in the region where polariton and phonon dispersion cross. In a two-dimensional system with parabolic dispersion relation the density of states is given by a step function with the amplitude given by the effective mass. The effective mass of the photon field is orders of magnitude lower than the effective mass of the exciton. Therefore the scattering probability in the low-energy region above the lowest LP state is strongly reduced due to the low density of states and essentially no dephasing due to resonant acoustic phonon absorption takes place. This leads to the constant FWHM of the LP in Fig.~\ref{Fig_det_specs_full}(a). The small, but finite dependence of the LP FWHM on the detuning in Fig.~\ref{Fig_det_specs_full}(a) results from the non-diagonal coupling of LP and UP such that a small contribution of the dephasing of the UP is added to the LP.

For increased temperatures, optical phonon absorption becomes important, because optical phonons carry a momentum-independent energy larger than the polariton binding energy and may thereby lead to a finite dephasing of the LP as observed in Fig.~\ref{Fig_det_specs_full}(b) for $T=150~$K. This explains the weaker dependence on detuning of the LP FWHM in Fig.~\ref{Fig_det_specs_full}, because mainly optical phonon scattering determines the respective linewidth.

We emphasize that this strict decoupling of the LP from acoustic phonons is specific to 2D materials where the phonon-mode itself is confined to the 2D plane. In, e.g., GaAs quantum wells the 2D carriers interact with 3D phonons such that acoustic phonons may overcome the energy-mismatch between the 2D-dispersion relations in Fig.~\ref{Fig_det_specs_full}(c) by adding a transverse momentum $K_z$ orthogonal to the quantum well. While it has been shown that also these processes do not lead to a strong dephasing process in a 2D quantum well \cite{Savona97}, it is completely absent in 2D materials. 

\subsection{Phonon side bands}

\begin{figure}[b]
\includegraphics[width=\columnwidth]{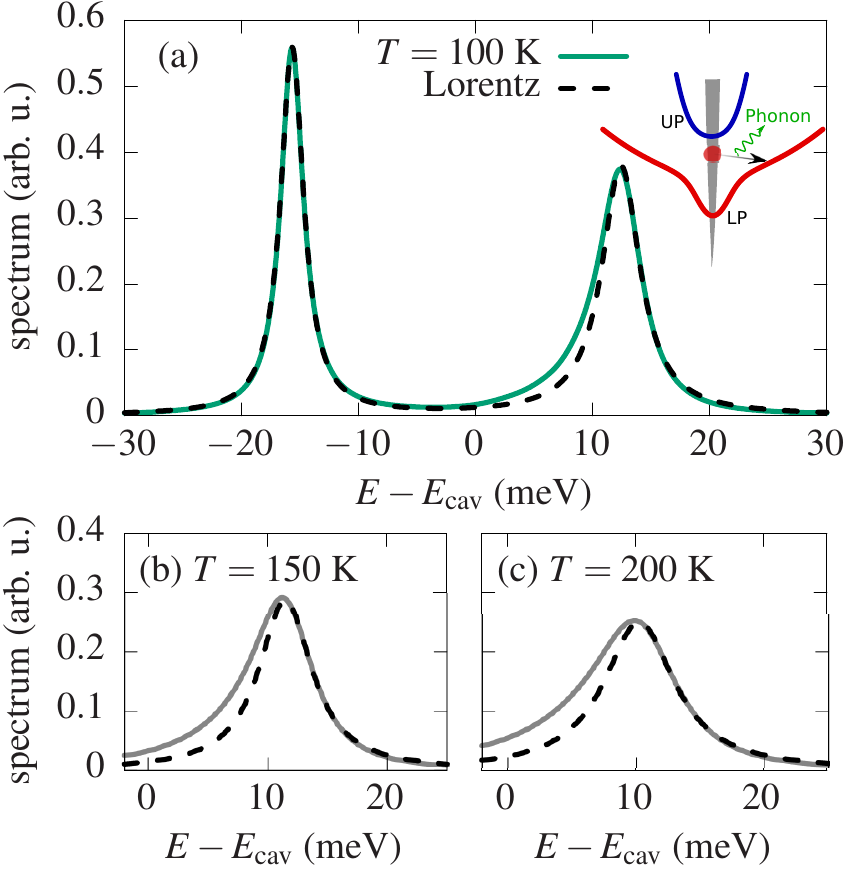}
\caption{Spectrum together with Lorentzian fits (black dashed lines) for (a) both polaritons and $T=100~$K and (b),(c) only for the UP for $T=150~$K and $T=200~$K, respectively. The inset in (a) sketches the phonon-assisted transitions which lead to the increased spectral weight below the UP resonance.}
\label{Fig_spectra_100K_Lorentz}
\end{figure}

Another remarkable feature of the polariton spectra is an asymmetry of the UP line for increased temperatures. As an example we plot the spectrum for $T=100~$K in Fig.~\ref{Fig_spectra_100K_Lorentz}(a) together with a sum of two Lorentzian fits adjusted to match the high-energy flank of the LP and UP resonance, respectively (black dashed line). Indeed, the UP resonance does not show a symmetric line, but has an increased spectral weight on the low-energy side, while the LP is well described by a Lorentzian. For increasing temperature this asymmetry of the UP gets more pronounced as shown in Fig.~\ref{Fig_spectra_100K_Lorentz}(b) and (c), where we plot the spectral range of the UP with a Lorentzian fit to the high-energy flank for $T=150~$K and $T=200~$K, respectively. For increased temperatures also the LP gains a small asymmetry, but the asymmetry of the UP is always dominant (not shown). Numerical tests show that the origin of this asymmetry is the coupling to acoustic phonons rather than optical phonons. The phonon-assisted processes leading to the low-energy flank of the UP are sketched in the inset of Fig.~\ref{Fig_spectra_100K_Lorentz}(a).

An optical transition of a phonon-assisted process may take place at energies $E=E_{\mathbf{K},\Lambda}\pm\hbar\omega_{\mathbf{K},i}$ by phonon emission ($+$) or absorption ($-$). Because acoustic phonons only transfer a small energy, the LP states with finite momentum $\mathbf{K}$ (red) below the UP (blue) may be reached after optical excitation of a virtual state (red circle within the grey light cone) by such a process (indicated by black arrows) and those transitions are the origin of the observed low-energy flank. 

Such a consideration would however also lead to a high-energy flank of the LP similar to the high-energy flank of the absorption spectrum as found for a bare monolayer on a substrate due to acoustic phonons \cite{Christiansen17,Shree18}. However, the peculiar dispersion relation of the LP leads to the absence of such processes. In the low-energy region of the LP, the density of states is vanishingly small as sketched in Fig.~\ref{Fig_det_specs_full}(c) and therefore the probability of a phonon-assisted transition into states directly above the lowest LP state is negligible. Therefore we do not observe an asymmetry of the LP in Fig.~\ref{Fig_spectra_100K_Lorentz}(a). We note however that for higher temperatures the optical phonons lead to a considerable coherent coupling of LP and UP such that also the LP gains a small amount of asymmetry, which is however in any case lower than for the UP.

\subsection{Influence of inhomogeneous broadening}
\label{Sec_Inhom}
Until now we have only considered homogeneous broadening mechanisms in the description of the polariton spectra. Monolayers of TMDCs usually show a pronounced inhomogeneous broadening of the exciton line due to the sensitivity on the surrounding material. Even though inhomogeneous broadening can be drastically reduced, mainly by means of hBN-encapsulation \cite{Cadiz17,Jakubczyk19}, it is instructive to include an inhomogeneously broadened exciton ensemble in the calculation. 


To account for the inhomogeneous broadening we introduce an ensemble of excitons with a varying exciton-cavity detuning $\Delta$ by defining the exciton energy as $E_{\mathbf{K}}(\Delta)=E_{\text{cav}}+\Delta+\frac{\hbar^2K^2}{2M}$. We assume that every detuning $\Delta$ couples independently to the photon modes and thereby
the exciton and therefore the polariton operators depend on the detuning with
	$$\hat{P}_{\mathbf{K},\Lambda}(\Delta)=\alpha_{\mathbf{K},\Lambda}(\Delta)\hat{X}_{\mathbf{K}}(\Delta)+\beta_{\mathbf{K},\Lambda}(\Delta)\hat{a}_{\mathbf{K}},$$
where the coefficients $\alpha,\beta$ depend on $\Delta$ [cf. eigenvectors in Eq.~\eqref{Eq_Polariton_Basis}]. The full spectrum is then calculated by integrating over all detunings
	$$\langle\tilde{\hat{a}}_{\mathbf{0}}\rangle(E)=\sum\limits_{\Lambda}\int \alpha_{0,\Lambda}(\Delta)\rho(\Delta)\langle\tilde{\hat{P}}_{\mathbf{0},\Lambda}\rangle(\Delta,E) d\Delta$$
where $\langle\tilde{\hat{P}}_{\mathbf{0},\Lambda}\rangle$ is the Fourier-transformed polariton amplitude. $\rho(\Delta)$ is the normalized distribution of detunings $\Delta$, which we assume to be Gaussian 
	$$\rho(\Delta)=\frac{1}{\sqrt{2\pi}D}\exp\left(-\frac{(\Delta-\Delta_0)^2}{2D^2}\right)$$
with the energetic width $D$ around the central detuning $\Delta_0$. We can regain the results of the former section by taking $D\rightarrow 0$.

\begin{figure}[tb]
\includegraphics[width=\columnwidth]{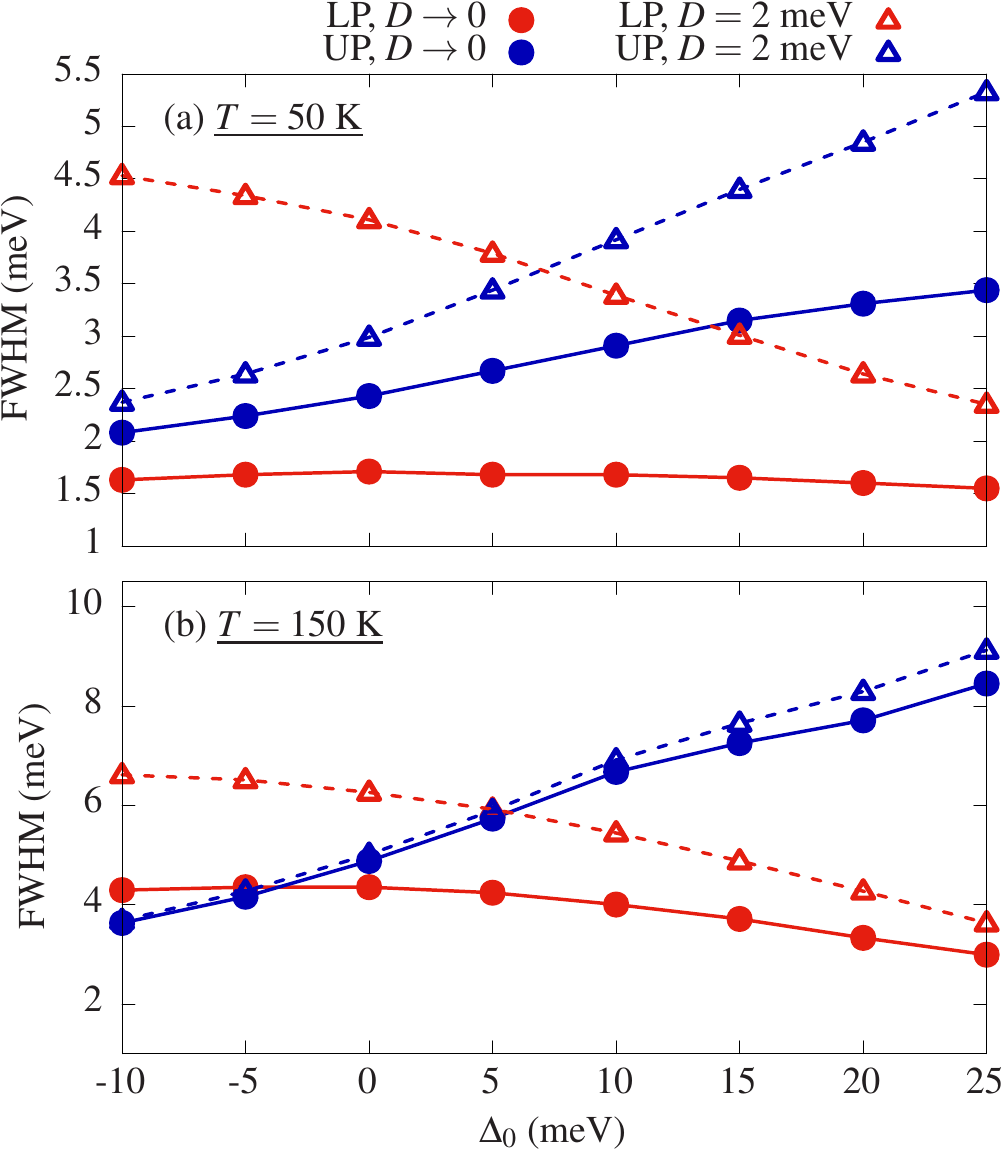}
\caption{FWHM of LP (red) and UP (blue) resonance for (a) $T=50~$K and (b) $T=150~$K. The filled circles (solid lines) show the case without inhomogeneous broadening and the empty triangles (dashed lines) show the results with an inhomogeneous broadening of $D=2~$meV.}
\label{Fig_det_specs_full_inhom}
\end{figure}

The inhomogeneous broadening mainly affects the polariton resonance with dominant exciton contribution, while a purely photonic mode is unaffected. This effect is shown in Fig.~\ref{Fig_det_specs_full_inhom}(a), where we plot the FWHM for $T=50~$K without inhomogeneous broadening (filled circles) [cf. Fig.~\ref{Fig_det_specs_full}(a)] for LP (red) and UP (blue) together with a calculation with a finite inhomogeneous broadening of $D=2~$meV (empty triangles). 
It is directly evident that the linewidth of the LP is strongly dominated by the influence of the inhomogeneous broadening, as it is essentially constant when only accounting for homogeneous broadening. For negative detuning $\Delta_0$ the excitonic component of the LP increases, which is therefore affected stronger by inhomogeneous broadening. In reverse, for positive detuning the LP becomes more photonic and accordingly the FWHM of the LP with inhomogeneous broadening approaches the calculation with $D\rightarrow 0$ (filled circles) for increasing $\Delta_0$. The impact of the inhomogeneous broadening on the UP line behaves in the opposite way: For negative detuning, the inhomogeneous broadening changes the FHWM only slighty, while for positive detuning the UP become more excitonic and a strong influence of the inhomogeneous broadening is seen. This finding is in agreement with Ref.~\cite{Dufferwiel15}, where a crossover from a broad UP to a broad LP signal was observed when tuning the detuning through resonance at $T=4.2~$K. 

When increasing the temperature to $T=150~$K in Fig.~\ref{Fig_det_specs_full_inhom}(b) the relative influence of the inhomogeneous broadening decreases because the homogeneous broadening by phonons dominates. The influence on the UP (blue) is barely visible because for strong negative detuning the UP is photon-like and therefore not coupled to the inhomogeneous broadening and for positive detunings the homogeneous broadening strongly dominates the FWHM. The influence on the LP is still larger because the influence of acoustic phonons is essentially absent as shown above, but the relative influence as compared to the case of $T=50~$K is reduced because of a finite contribution due to optical phonon absorption. 

\section{Conclusions}
In conclusion we performed non-Markovian calculations of the phonon-induced line shape of polariton spectra of a $\mathrm{MoSe_2}$ monolayer embedded in a high-Q microcavity by means of a time convolutionless master equation. The temperature-dependent detuning of the exciton and cavity influences the effective photon contribution of the upper and lower polaritons and therefore their general visibility in the spectra. We found a non-trivial dependence of the spectral linewidths of both resonances as function of temperature and detuning due to the influence of phonons. The peculiar polariton dispersion relation leads to a LP linewidth that is largely independent of the acoustic phonon contributions. In contrast, acoustic phonons lead to a low-energy shoulder of the UP resonance in the spectra due to phonon-assisted processes involving the exciton-like part of the LP branch. We further discussed the impact of inhomogeneous broadening of the exciton ensemble on the spectra and showed that this affects especially the LP linewidth at low temperatures.
 
The presented results contribute to a thorough understanding of polariton-phonon interaction and will help to interpret and optimize the optical properties of TMDC-based polaritonic devices. 

\acknowledgements
 F.~L. and D.~E.~R. acknowledge financial support by the Deutsche Forschungsgemeinschaft (DFG) by the project
406251889 (RE 4183/2-1). 
\appendix

\section{Derivation of the TCL equation}
\label{App_Derivation}
We here give a brief derivation of the equation of motion of the polariton amplitude $P_{\mathbf{K},\Lambda}$ from Eq.~\eqref{Eq_EOM} starting from the Hamiltonian in Eq.~\eqref{Eq_H_PolaritonBasis}. Starting from the Ehrenfest theorem, the coherent polariton amplitude $P_{\mathbf{K},\Lambda}$ couples to the phonon-assisted density matrices $S_{\mathbf{K},\mathbf{Q},\Lambda,i}^{(-)}:=\langle\hat{P}_{\mathbf{K-Q},\Lambda}\hat{b}_{\mathbf{Q},i}\rangle$ and $S_{\mathbf{K},\mathbf{Q},\Lambda,i}^{(+)}:=\langle\hat{P}_{\mathbf{K-Q},\Lambda}\cre{b}_{-\mathbf{Q},i}\rangle$  according to 
	\begin{align*}
	\frac{d}{dt}P_{\mathbf{K},\Lambda}=
	&-\frac{i}{\hbar}\mathcal{E}_{\mathbf{K},\Lambda}P_{\mathbf{K},\Lambda}\\
	&-\frac{i}{\hbar}\sum\limits_{\mathbf{Q},i,\Lambda''}g_{\mathbf{Q},\mathbf{K-Q},i}^{\Lambda\Lambda''}\left(S_{\mathbf{K},\mathbf{Q},\Lambda'',i}^{(-)}+S_{\mathbf{K},\mathbf{Q},\Lambda'',i}^{(+)}\right).
	\end{align*}
	
The equation of motion for $S^{(\pm)}$ reads in second Born approximation \cite{Rossi02}
	\begin{align*}
	\frac{d}{dt}S_{\mathbf{K},\mathbf{Q},\Lambda'',i}^{(\pm)}=
	&-\frac{i}{\hbar}\left(\mathcal{E}_{\mathbf{K-Q},\Lambda''}\mp \hbar\omega_{\pm\mathbf{Q},i}\right)S_{\mathbf{K},\mathbf{Q},\Lambda'',i}^{(\pm)}\\
	&-\frac{i}{\hbar}\sum\limits_{\Lambda'}g_{-\mathbf{Q},\mathbf{K},i}^{\Lambda''\Lambda'}P_{\mathbf{K},\Lambda'}\left(\frac{1}{2}\mp\frac{1}{2}+n_{\mp\mathbf{Q},i}\right).
	\end{align*}

The TCL equation now results from a formal integration of $S^{(\pm)}$, where in the integrand $P_{\mathbf{K},\Lambda'}$ is approximated in zeroth Born approximation \cite{Lengers20} by
	$$P_{\mathbf{K},\Lambda'}(t-\tau)\approx e^{\frac{i}{\hbar}\mathcal{E}_{\mathbf{K},\Lambda'}\tau}P_{\mathbf{K},\Lambda'}(t)$$
in the spirit of a slowly-varying dynamics in the interaction picture.


%

\end{document}